\documentclass[twocolumn,amsmath,amssymb,showpacs,superscriptaddress,prb,aps,longbibliography,10pt]{revtex4-1}
\usepackage{graphicx}
\usepackage{bm}
\usepackage{dcolumn}
\usepackage{amssymb}
\usepackage{amsmath}
\usepackage{amsfonts}
\usepackage{epsfig}
\usepackage{graphicx}
\usepackage{stackrel}
\usepackage{paralist}
\usepackage[rgb]{xcolor}
\usepackage{todonotes}
\usepackage{pdfcomment}
\usepackage{mathrsfs}

\newcommand{\refEq}[1]{Eq.~(\ref{#1})}
\newcommand{\refFig}[1]{Fig.~\ref{#1}}

\newcommand{\citeRef}[1]{Ref.~\onlinecite{#1}}

\begin{document}
\title{Non-Local Coulomb Drag in Weyl Semimetals}
\author{Yuval Baum}
\affiliation{Department of Condensed Matter Physics, Weizmann Institute of Science, Rehovot 76100, Israel}
\affiliation{Institute of Quantum Information and Matter, Department of Physics, California Institute of Technology, Pasadena, California 91125, USA}
\author{Ady Stern}
\affiliation{Department of Condensed Matter Physics, Weizmann Institute of Science, Rehovot 76100, Israel}
\begin{abstract}
Non-locality is one of the most striking signatures of the topological nature of Weyl semimetals.  
We propose to probe the non-locality in these materials via a measurement of a magnetic field dependent Coulomb drag between two sheets of graphene which are separated by a three-dimensional slab of Weyl semimetal. We predict a new mechanism of Coulomb drag, based on cyclotron orbits that are split between opposite surfaces of the semi-metal. In the absence of impurity scattering between different Weyl nodes, this mechanism does not decay with the thickness of the semi-metal. 
\end{abstract}
\maketitle

\section{Introduction}
Weyl semimetals (WSM) are a new class of conducting materials, characterized by the nontrivial topological structure of their band structures. These three-dimensional topological semimetals have been the subject of intense recent theoretical and experimental efforts \cite{TurnerRev,Turner2,Burkov,Hosur,Material1,Material2,Bernevig,Exp1,WeylExp1,WeylExp2,Exp2,TaAs2}.

WSM host an even number of points in the Brillouin zone, known as the Weyl nodes, at
which the bulk energy gap closes. The electrons around these nodes disperse relativistically, and may be described at low energies by the Weyl Hamiltonian \cite{TurnerRev,Turner2,Burkov}.
Each Weyl node acts as either a source or a sink of Berry flux. Hence, a single node is stable and cannot be removed. Additionally, any closed two-dimensional manifold in the three-dimensional Brillouin zone (BZ) which separates regions with different Berry charge must have a non-zero Chern number as it is threaded by a non-zero Berry flux. Thus, as long as Weyl nodes of opposite Berry charge, are separated in momentum space, these systems host 2D surface states which form Fermi-arcs in the two-dimensional BZ of the surface \cite{Turner2}.

Certain iridium pyrochlores and noncentrosymmetric transition metals, e.g. TaAs, have been predicted to host such a topological semimetal phase \cite{Material1,Material2,Bernevig,Exp1}. Following these suggestions, several topological semimetals have been observed experimentally \cite{WeylExp1,WeylExp2,WeylExp3,WeylExp4,WeylExp5,Exp1,Exp2,TaAs2,TaAs3,TaAs4,Batabyale,Inoue}.

It has been proposed that, when subject to a magnetic field, such materials exhibit unusual physical phenomena, such as
the Adler-Bell-Jackiw chiral anomaly \cite{chiralAnSpivak,ciralAnExp,chiralexp2,chiralexp2}, unique type of quantum oscillations associated with Fermi-arc surface states \cite{Potter1, Moll} and various non-local transport effects \cite{nonlocalSid,Baum16}.
In particular, \citeRef{Baum16} proposed two experiments that directly probe the exotic topology of these materials: appearance of a magnetic-field-dependent non-local dc voltage and sharp resonances in the transmission of electromagnetic waves. Both these effects do not rely on quantum mechanical phase coherence, which renders them less restrictive in terms of temperature and samples mobility. 
Nevertheless, they do require challenging experimental setups such as placing contacts in a close proximity in the dc proposal and working with a THz frequency radiation in the ac proposal.

In this paper, we propose an alternative dc experiment that does not require any deposition of contacts over the Weyl semimetal: a magnetic-field-dependent non-local Coulomb drag measurement.  
The underlying mechanism in this work is similar to one presented in \citeRef{Baum16,Potter1,nonlocalSid}. When a magnetic field is applied perpendicular to surfaces that carry Fermi arcs, electrons traverse cyclotron orbits that connect opposite surfaces of the sample. These unusual inter-surface orbits are a source of non-local conductivity in these materials, i.e., the application of an electric field along a surface of these materials generates a current along the opposite surface and vice versa. 
In the following, we provide a prescription which is aimed to detect the emergence of non-locality via a Coulomb drag experiment. 

Standard Coulomb drag setups involve two closely spaced, yet electrically isolated conductors.
Due to the long-range Coulomb interaction between charge carriers in the two conductors, a current in one of the conductors (the active layer) induces a voltage (or a current) in the other conductor (the passive layer). The ratio between the induced voltage in the passive layer and incoming current in the active layer is the drag resistivity \cite{DragRev1,DragRev}. 

Drag measurements may be used to study fundamental properties related to the e-e interaction of diverse condensed matter systems. 
Until recently, semiconductor heterostructures, e.g. GaAlAs, with two quantum wells separated by a thin tunnel barrier were the main experimental systems allowing
reliable measurements of Coulomb drag \cite{DragGaAs1,DragGaAs2,DragGaAs3}. The dependence of the drag effect on the layer densities, temperature, and separation
distance between the quantum wells has been heavily investigated. In these experiments, the typical separation between the layers was in the range $15-50\mbox{ nm}$.
In recent years, the appearance of graphene-boron-nitride heterostructures, made it possible to investigate drag effects in double-layer graphene down to a separation of $\sim 1\mbox{ nm}$ \cite{GrapDragGeim,DragGrphTh1,DragGrphTh2,DragGrphTh3,DragGrphTh4}.
 
In this work, we consider a scenario where two layers of graphene are separated by a 3D slab of WSM. The graphene layers are coupled to the WSM only via the Coulomb interaction, i.e., the top and bottom graphene layers generate drag force on the top and bottom surfaces of the WSM respectively. In the absence of an external magnetic field the system may be thought of as two decoupled standard Coulomb drag systems, each one contains a layer of graphene which is Coulomb coupled to a surface of a WSM. Since the WSM is metallic, the coupling between these two systems is non-zero, nonetheless, for macroscopic slabs this coupling is extremely small as we show below.
On the contrary, in the presence of an external magnetic field, the non-locality in the WSM yields a strong coupling between the two drag systems and a current in the top graphene layer leads to a non-negligible voltage buildup on the distant lower graphene layer.

The source of the non-local drag is rather simple. As long as current flows in the top graphene layer, a portion of its momentum is transferred to the top surface of the WSM, due to the existence of a frictional Coulomb force. For a standard 2DEG in an open circuit configuration, this momentum transfer is translated into a voltage buildup. However, as shown in \citeRef{Baum16,Potter1}, for relatively clean samples and in the presence of an external magnetic field the current on the top surface follows the inter-surface cyclotron orbits leading to a circulating current throughout the 3D sample. In particular, a current on the bottom surface is inevitable for clean samples. When the non-local conductivity is much larger than the local one, the current on the bottom surface flows with a similar magnitude and an opposite direction to the current on the top surface.
As before, due to the existence of a frictional Coulomb force between the bottom surface of the WSM and the bottom graphene layer, a momentum transfer to the second graphene layer occurs. This momentum transfer is translated into a voltage buildup on the passive graphene layer. The ratio between this voltage and the current on the top layer is the non-local drag coefficient.
 
\section{Setup and transport equations}
\begin{figure}[t!] 
\begin{center}
{\includegraphics[width = \linewidth, angle =0] {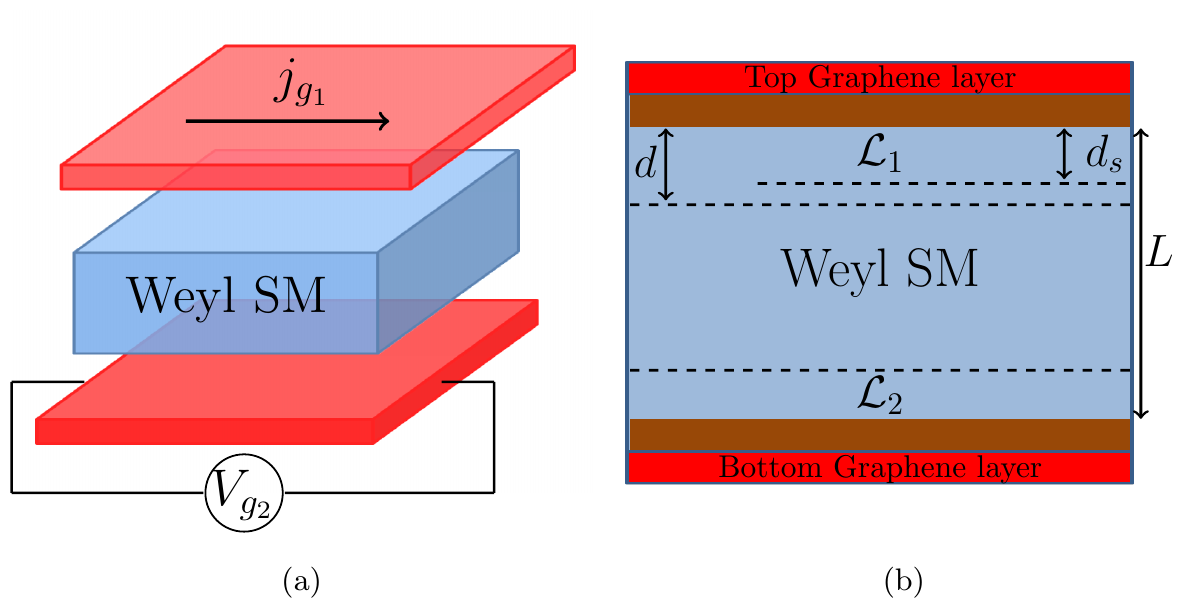}}
\end{center}
\caption{(a) Drag setup: two layers of graphene (red) are placed above and below a 3D Weyl semimetal. The graphene layers and the Weyl semimetal are separated by insulating layers (brown). The graphene layers are coupled to the Weyl metal only via the local Coulomb interaction. Current is pushed on the top layer while the voltage on the bottom layer is measured.
(b) A sketch of the relevant length scales in the problem. $L$ is the thickness of the WSM, $d_s$ is the extent of the surface states and $d$ is the thickness of the layer at which the drag coupling is non-zero. We denote the layers at which the surface states exist by $\mathcal{L}_1$ and $\mathcal{L}_2$.}\label{weyl_sys}
\end{figure}

In this section, we describe the experimental setup and the procedure of finding the non-local drag coefficient. The setup we consider is depicted in \refFig{weyl_sys}a. Two layers of graphene (red) are placed above and below a 3D Weyl metal. The graphene layers are coupled to the WSM only via the Coulomb interaction, i.e., the tunneling between the graphene layers and the WSM is assumed to be zero. A current $j_{g_1}$ is pushed in the first graphene layer while the voltage (electric field) buildup on the second graphene layer, $V_{g_2}\,(\epsilon_{g_2})$, is measured. The ratio between these two quantities, $R_D=\epsilon_{g_2}/j_{g_1}$, is the non-local drag coefficient. 

We analyze the semi-classical transport equations, assuming that the relevant transport parameters are known. For simplicity, we assume that the two graphene layers are identical. We denote their 2D density by $n_g$ and their 2D intra-layer resistivity by $\rho_g$. Moreover, at this stage, we assume that the Coulomb coupling between the WSM and the two graphene layers is symmetric and can be characterized by a single 2D drag resistivity $\rho_d$. The form of $\rho_d$ is given in appendix \ref{drag_app}. Finally, we denote the 3D density and the 3D resistivity of the WSM by $n_{3D}$ and $\rho_{3D}$ respectively.

A standard double-layer Coulomb drag system is well described by two coupled Drude-like equations of motion,
\begin{align} \label{stand_drag}
&m_1\frac{d\textbf{v}_1}{dt}=e\left(\textbf{E}_1+\textbf{v}_1\times\textbf{B}\right)-\frac{m_1\textbf{v}_1}{\tau_1}-\hbar\gamma n_2(\textbf{v}_1-\textbf{v}_2)\\
&m_2\frac{d\textbf{v}_2}{dt}=e\left(\textbf{E}_2+\textbf{v}_2\times\textbf{B}\right)-\frac{m_2\textbf{v}_2}{\tau_2}-\hbar\gamma n_1(\textbf{v}_2-\textbf{v}_1) \nonumber
\end{align}
where $\textbf{B}$ is the magnetic field, $e$ is the electric charge, $\textbf{v}_i$, $m_i$, $n_i$ and $\textbf{E}_i$ are the drift velocity, effective mass, density and electric field in the $i$'s layer, respectively. $\tau_i$ is the $i$'s layer intra-layer momentum-relaxation-time due to impurity-scattering while the dimensionless parameter $\gamma$ is related to the mutual inter-layer momentum-relaxation-rate due to the Coulomb friction. The drag resistivity arising from \refEq{stand_drag} is $\hbar\gamma/e^2$. In the following, we assume steady state, i.e., the left hand size above is zero.

In the setup we propose, each graphene layer is Coulomb coupled locally to the adjacent surface of the 3D WSM, and the WSM, similar to \citeRef{Baum16}, has both a local and a non-local conductivity.
We denote the three-dimensional electric field and the current density in the WSM by $\textbf{E}(x,y,z)$ and $\textbf{J}(x,y,z)$. 
The current density and electric field in the WSM must satisfy Kirchoff's rules and Ohm's law, 
\begin{align}
 \label{kirch}
\nabla\cdot\textbf{J}&=\nabla\times\textbf{E}=0 \\
\textbf{E}&=\hat{\rho}\textbf{J} \nonumber
\end{align}
where the resistivity matrix includes both the local and the non-local parts.
In the limit where the thickness of the WSM (extent in the $z$-direction) is much smaller than its extent in the other directions, we may assume that the bulk values of $\textbf{E}$ and $\textbf{J}$ do not depend on $x$ and $y$. Hence, the requirement $\nabla\cdot \textbf{J}=0$ implies that $\partial_z J_z=0$ and therefore $J_z=0$, since it must be zero on the boundaries. Moreover, since the WSM is in an open circuit configuration, the net current in the 
3D bulk must be zero, i.e.,
\begin{align} \label{current} 
\int{dzJ_x(z)}=\int{dzJ_y(z)}=0
\end{align} 
where the integration is over the thickness of the WSM slab. Additionally, the requirement $\nabla\times \textbf{E}=0$ implies that $\textbf{E}=\mbox{const}$.

Next, we denote the two-dimensional electric fields and the current densities in the $i$'s graphene layer by $\textbf{j}_{g_i}(x,y)$ and $\boldsymbol {\epsilon}_{g_i}(x,y)$. The currents in the graphene layers are taken to be controlled quantities in the experiment while the spatially independent electric fields are the unknowns which should be determined.
 
The above analysis points out that we are left with five unknown quantities, three uniform fields $\boldsymbol {\epsilon}_{g_1},\,\boldsymbol {\epsilon}_{g_2},\,\textbf{E}$ and two $z$-dependent currents $\textbf{J}_x(z)$ and $\textbf{J}_y(z)$.
In order to determine these unknown quantities, we invoke Ohm's law for the combined graphene-WSM system. 
In appendix \ref{Hall_app} we show that the effect of the intra-layer Hall effect on the non-local drag is negligible. Hence, we ignore the Lorentz force in \refEq{stand_drag}.

The electric field in the graphene layers is determined only by the intra-layer scattering and the local Coulomb drag:
\begin{align} \label{ohm_differential1} 
&\boldsymbol {\epsilon}_{g_1}=\left(\rho_g+\frac{n_{3D}}{n_{g}}\int dz'\rho_{d1}(z')\right)\textbf{j}_{g_1}-\int dz'\rho_{d1}(z')\textbf{J}(z')\\ \nonumber
&\boldsymbol {\epsilon}_{g_2}=\left(\rho_g+\frac{n_{3D}}{n_{g}}\int dz'\rho_{d2}(z')\right)\textbf{j}_{g_2}-\int dz'\rho_{d2}(z')\textbf{J}(z')
\end{align}
where $\rho_{di}(z)$ is the local drag resistivity that couples the $i$'s graphene layer to a fixed $z$ layer of the WSM, where the integrals run over the thickness of the sample. Both $\rho_{d1}$ and $\rho_{d2}$ are local quantities, and hence, vanish away from the top and bottom surfaces respectively.
Since $\textbf{j}_{g_1}/(en_g)$ is the velocity in the graphene layer and $\textbf{J}(z)/(en_{3D})$ is the velocity in the WSM, it is evident, as expected, that when the velocities are equal the drag is zero.

In the WSM, there are three contributions to the resistivity: the impurity scattering, Coulomb drag with both graphene layers and an additional contribution coming from the inter-surface cyclotron orbits as derived in \citeRef{Baum16}. Overall, Ohm's law in the WSM is given by, 
\begin{align} \label{ohm_differential2} 
&\textbf{E}=\left(\rho_{3D}+\frac{n_{g}}{n_{3D}}\left[\rho_{d1}(z)+\rho_{d2}(z)\right]\right)\textbf{J}(z)\\ \nonumber
&\,\,\,\,\,-\rho_{d1}(z)\textbf{j}_{g1} 
-\rho_{d2}(z)\textbf{j}_{g2}+\int dz'\rho_{w}(z,z')\textbf{J}(z') \\ \nonumber 
\end{align}
where $\rho_{w}(z,z')$ is the contribution of the inter-surface cyclotron orbits to the resistivity of the WSM. In general, $\rho_{w}(z,z')$ contains both local and non-local contributions.

For concreteness, we denote the thickness of the WSM by $L$. The Coulomb interaction between the graphene layers and the WSM bulk vanishes exponentially away from the surface, where the decaying scale is determined by the Thomas-Fermi screening length. Therefore, the coupling between the graphene layers and the WSM is highly local. For simplicity, we assume that the coupling is uniform inside thin layers near the surfaces of the WSM and zero outside of these layers, i.e.,
\begin{align} \label{drag_struc} 
\rho_{d1}=
\begin{cases}
\rho_d,\,\, 0<z<d\\
0, \,\,\mbox{otherwise}
\end{cases},\,\rho_{d2}=
\begin{cases}
\rho_d,\,\, L-d<z<L\\
0, \,\,\mbox{otherwise}
\end{cases}
\end{align}
where we denoted the thickness of the layers in which the drag is effective by $d$. We assume $d\ll L$. In the presence of an external magnetic field, non-local effects connect the top and bottom surface of the WSM. We denote the extent of the surface states by $d_s \ll L$. All these scales are depicted in \refFig{weyl_sys}b. With these assumptions, we may write $\rho_{w}(z,z')$ as:

\begin{align} \label{non_loc_struc} 
\frac{\rho_w(z,z')}{d_s\rho_{3D}}=
\begin{cases}
\,\,\,\,\,\Gamma,\,\, z,z'\in \mathcal{L}_1\\
\,\,\,\,\,\Gamma,\,\, z,z'\in \mathcal{L}_2\\
-\Gamma, \,\,z\in \mathcal{L}_1,\,z'\in \mathcal{L}_2\\
-\Gamma, \,\,z\in \mathcal{L}_2,\,z'\in \mathcal{L}_1\\
\end{cases}
\end{align}
where $\mathcal{L}_1$ is the layer defined by $z\in [0,d_s]$, $\mathcal{L}_2$ is the layer defined by $z\in [L-d_s,L]$ and the dimensionless parameter $\Gamma$ is specified below.

In order to solve \refEq{ohm_differential1} and \refEq{ohm_differential2} with the appropriate boundary conditions and the constraint of \refEq{current}, we discretize the 3D WSM to $N$ quasi-2D layers. The full description of the solution appears in appendix \ref{dis_sol}.

Before diving into the details of the exact solution, it is instructive to qualitatively examine the case without non-locality which is applicable when no magnetic field is present. In this case, a current in the top graphene layer leads to a current, in the same direction, 
in the top layer of thickness $d$ of the WSM. For weak drag this current is given by $j_{top}=(\rho_d j_{g_1})/\rho_{3D}$.
 Away from the top surface \refEq{ohm_differential2} implies that the current in the bulk is uniform. However, \refEq{current} requires that the total current is zero. Hence an opposite current must flow in the bulk of the WSM in order to cancel the current in the top drag layer. The magnitude of this current must scale as $d/L$. In turn, this opposite current leads to voltage buildup on the second graphene layer due to its Coulomb coupling. The electric field in the bottom graphene layer is determined by the current in the bottom drag layer of the WSM, $\epsilon_{g_2}=\rho_d j_{bot}$. Finally by using the following currents ratio $j_{bot}/j_{top}= d/L$ we expect 
$R_D\approx\left(\frac{d}{L}\right)\frac{\rho_d^2}{\rho_{3D}/d}$. In the next section we show that the effect of non-locality is to increase the current flow near the bottom surface and hence increasing the non-local drag resistivity. 
 
\section{The Non-local Drag coefficient}
We express the non local drag coefficient by the three length scales above and the following quantities: the local sheet resistivity $r_0=\rho_{3D}/L$, the dimensionless non-local parameter $\Gamma$ and the effective density ratio $\xi=\frac{n_{3D}L}{n_g}$ which generically is much larger than unity.
Solving for the non-local drag coefficient $R_D=\epsilon_{g_2}/j_{g_1}$ and employing the above notations yields the following lengthy expression whose limits we aim to examine:
\begin{equation} 
R_D=\frac{d_s\xi^2r_0\rho_d^2\left(\frac{d}{d_s}(\rho_d+\xi r_0)+\Gamma\left[A\rho_d+B\xi r_0\right]\right)}{L(\rho_d+\xi r_0)\left[A\rho_d+\frac{L}{d}\xi r_0\right]\left[\rho_d+\xi r_0(1-2\Gamma)\right]} \label{NLD}
\end{equation} 
where $A=\frac{L}{d}-2$ and $B=\frac{L}{d}-\frac{2d}{d_s}$.

In the following parts, we investigate the non-local drag coefficient in the relevant limits. 
For macroscopic samples we may assume $\frac{L}{d}\gg1$ and $\rho_d\ll\xi r_0$. In this limit \refEq{NLD} becomes:
\begin{equation} 
R_D\approx\frac{dd_s\xi\rho_d^2}{L^2}\frac{\frac{d}{d_s}+\Gamma\frac{L}{d}}{\rho_d+\xi r_0(1-2\Gamma)} \label{NLD1}
\end{equation} 

In the absence of an external magnetic field and in the limit of weak drag, \refEq{NLD} reduces to the expected result, as discussed in the previous section.

We now turn to the more interesting case in which $\Gamma\neq 0$, i.e., in the presence of an external magnetic field. It is constructive to introduce the notation,
\begin{equation} 
r_0(1-2\Gamma)=\frac{r_0r_w}{2r_0+r_w},
\end{equation}
with $Lr_w=1/\sigma_w$, where $\sigma_w$ is the non-local conductivity as derived in \citeRef{Baum16}. In the local case $r_w$ diverge to infinity and therefore $\Gamma=0$, while when the non-local conductivity is much larger than the local one then $r_w\ll r_0$ and therefore $\Gamma\to 1/2$.
Employing the above notation, the non-local drag coefficient is given by,
\begin{equation} 
R_D\approx \frac{\xi^2r_0\rho_d^2\left[\left(\frac{d}{L}\right)^2+\Gamma\left(\frac{d}{L}\right)\right]}{(\rho_d+\xi r_0)\left[\rho_d+\xi \frac{r_0r_w}{2r_0+r_w}\right]}
\approx\frac{\Gamma n_{3D}d_s}{n_g}\frac{\rho_d^2}{\rho_d+\xi \frac{r_0r_w}{2r_0+r_w}} \label{NLRD}
\end{equation} 

For low magnetic fields, the non-local conductivity is small compared to the local one, i.e., $r_w\gg r_0$. In this limit and assuming as before $\rho_d\ll\xi r_0$, \refEq{NLRD} yields:
\begin{equation} 
R_D\approx \Gamma \frac{d_s}{L}\frac{\rho_d^2}{r_0}= \Gamma \frac{\rho_d^2}{\rho_{3D}/d_s}+\mathcal{O}\left(\frac{d}{L}\right) \label{NL_w}
\end{equation} 

Remarkably, we find that in the presence of an external magnetic field the leading term in $R_D$ becomes independent of the WSM thickness $L$, consensuses of the inter-surface cyclotron orbits. This result is valid as long as the circulating inter-surface current exists. As shown in \citeRef{Baum16,Potter1}, this is the case whenever the thickness of the WSM is smaller than the inter-node-transport-scattering-length. In these materials it is expected  that the inter-node scattering length should be much smaller than the single-particle scattering length \cite{WeylNonLocalTransport}.

If the magnetic field is further increased, the non-local conductivity becomes larger than the local one, i.e., $r_w\ll r_0$ and $\Gamma \to \frac{1}{2}$. In this limit \refEq{NLRD} becomes:
\begin{equation} 
R_D\approx \frac{n_{3D}d_s}{n_g}\frac{\rho_d^2}{2\rho_d+\xi r_w}+\mathcal{O}\left(\frac{d}{L}\right)
\end{equation} 
which may be split into two cases:
for $ \rho_d\ll\xi r_w$,
\begin{equation} 
R_D\approx \frac{d_s}{L}\frac{\rho_d^2}{r_w}+\mathcal{O}\left(\frac{d}{L}\right)
\end{equation} 
which is similar to \refEq{NL_w} only with $\Gamma/r_0$ being replaced by $1/r_w$. Remember that in this limit $r_w\ll r_0/\Gamma$.

On the other hand, if $r_w$ is further reduced such that $\xi r_w\ll \rho_d$ then,
\begin{equation}
R_D\approx \frac{n_{3D}d_s}{n_g}\frac{\rho_d}{2}+\mathcal{O}\left(\frac{d}{L}\right) \label{G2}
\end{equation}

Surprisingly, in this case $R_D\propto \rho_d$. It is instructive to reproduce \refEq{G2} in the non-symmetric case where the drag resistivity on the top and the drag resistivity on the bottom are different. We denote the different local drag resistivities by $\rho_{d1}$ and $\rho_{d2}$ respectively. In the non-symmetric case \refEq{G2} becomes:
\begin{equation} 
R_D\approx \frac{n_{3D}d_s}{n_g}\frac{\rho_{d_1}\rho_{d_2}}{\rho_{d_1}+\rho_{d_2}}=\frac{n_{3D}d_s}{n_g}\left[\frac{1}{\rho_{d_1}}+\frac{1}{\rho_{d_2}}\right]^{-1} \label{G3}
\end{equation}
which implies that $R_D$ is dominated by the minimal local drag resistivity. It worth noting that the non-local drag coefficient in \refEq{G2} may be larger than the local drag resistivity.

\section{Estimate of scales}
We may consider both Dirac metals (e.g. CdAs) and Weyl metals (e.g. TaAs, TaP, NbAs, NbP). In these materials, the typical bulk density is in the range $n_{3D}=3\cdot 10^{18}-3\cdot 10^{19}\,\mbox{cm}^{-3}$, while a typical density of graphene sheets is in the range $n_g=10^{11}- 10^{12}\,\mbox{cm}^{-2}$. The mobilities of WSM may vary dramatically between the different materials \cite{WeylExp1,mobility1,mobility2,mobility3}, see table~\ref{mobilities}.   
\begin{table}[b]
\centering
\resizebox{0.75\linewidth}{!}{
\begin{tabular}{c|c|c|c}
           & CdAs & NbP & TaAs \\
\hline
$T<1.5K$   & $9\cdot 10^6$ &  $5\cdot 10^6$ &  $5\cdot 10^5$  \\
$T\sim 130K$  & $4.2\cdot 10^4$ & $7\cdot 10^3$ & $1.5\cdot 10^3$   \\
$T\sim 300K$ & $1.5\cdot 10^4$ & $3\cdot 10^2$ & -  \\
\end{tabular}}
\caption{Typical mobilities (of thin films) in units of $\mbox{V}^{-1}\mbox{sec}^{-1}\mbox{cm}^{2}$. }\label{mobilities}
\end{table}

As a result, at moderate temperatures ($20K<T<150K$) the typical 3D resistivity of these materials is in the range $\rho_{3D}=5-100\,\mu\Omega\mbox{cm}$.

\begin{figure}[t]
\begin{center}
\centering
{\includegraphics[width = 1\linewidth, angle =0] {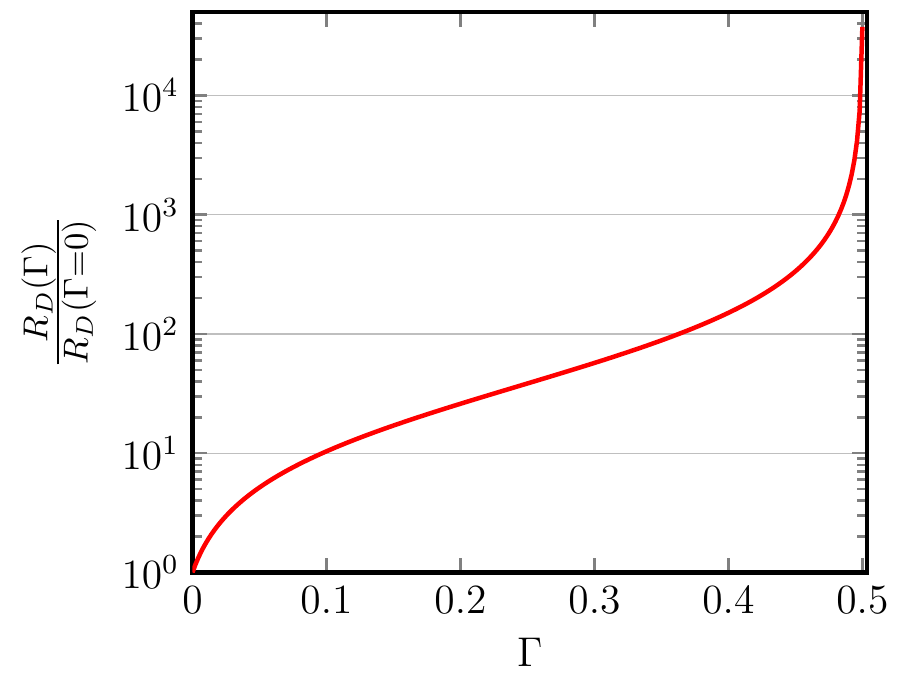}}
\end{center}
\caption{The ratio of $R_D$ in the presence of non-locality ($\Gamma\neq 0$) and $R_D$ in the absence of non-locality ($\Gamma= 0$). Here, $L=100\,\mbox{nm}$, $d_s=3\,\mbox{nm}$, $d=2\,\mbox{nm}$, $r_0=5\,\Omega$, $\rho_d= 0.5\,\Omega$ and $\xi=100$. For $\Gamma=0$ the drag coefficient is $R_D=2\cdot10^{-5}\,\Omega$ and for $\Gamma=1/2$ it is $R_D=0.75\,\Omega$.\label{RD_G}}
\end{figure}

With these parameters in hand we may estimate the relevant transport parameters of our system.
We consider a thin sample of a WSM with $L\sim100\,\mbox{nm}$, where both $d$ and $d_s$ are in the nanometer scale ($\sim 1-5\,\mbox{nm}$). Moreover, $r_0=\frac{\rho_{3D}}{L}=0.5-10\,\Omega$ and $r_0\xi=(en_g\mu)^{-1}=30-10^3\,\Omega$, where $\mu$ is the mobility of the WSM. At moderate temperatures, typical values for the local drag resistivity in graphene systems (away from the dual neutrality point) are in the range $\rho_d\sim 0.1-10\,\Omega$. \cite{GrapDragGeim} See appendix \ref{drag_app} for an estimation of $\rho_d$.

Employing these parameters, the ratio of $R_D$ in the presence of non-locality ($\Gamma\neq 0$) and $R_D$ in the absence of non-locality ($\Gamma= 0$) appears in \refFig{RD_G}. 
As evident, the non-local drag coefficient increases by orders of magnitude as a function of $\Gamma$. In particular, while the value of $R_D$ for $\Gamma=0$ is practically unmeasurable (of the order of $\mu\Omega$), it becomes of the order of $1\Omega$ for $\Gamma\sim 1/2$.

\section{Conclusions}
We have shown that the non-local conductivity in WSM which arises from inter-surface cyclotron orbits may be detected in a Coulomb drag measurement.
We demonstrated that, when the non-local conductivity is non zero, a measurable drag resistance is expected between two layers of graphene which are separated by hundreds of nanometers.
Such an unprecedented strong signal of drag in these scales provides a clear-cut evidence to the existence of non-local effects, and therefore, to the special topological structure of these materials. 

While in WSM with $C_4$ symmetry, the non-local drag is expected to be isotropic, in WSM which break the $C_4$ symmetry, the non-local drag is expected to be highly non-isotropic which is a direct measure to the unique arc structure of the surface states.  
Furthermore, similar to \citeRef{Baum16}, the semi-classical origin of this effect renders it more robust and experimentally accessible than quantum effects.

It is also worth mentioning that it may be possible to use the drag setup only along the top surface of the Weyl semimetal, while the detection of the non-local current may be done by a direct measurement of the voltage buildup on the bottom surface of the Weyl semimetal.

\begin{acknowledgments}
We thank Jim Eisenstein and Yuval Oreg for useful discussions. We also acknowledge support from the European Research Council under the European Unions Seventh Framework Program (FP7/2007-2013) / ERC Project MUNATOP, the DFG
(CRC/Transregio 183, EI 519/7-1), Minerva foundation, and the U.S.-Israel BSF. YB is grateful for support from the Institute of Quantum Information and Matter.
\end{acknowledgments}

\appendix
\section{Solution of the transport equations}\label{dis_sol}
In order to solve \refEq{ohm_differential1} and \refEq{ohm_differential2} with the appropriate boundary conditions and the constraint of \refEq{current}, we discretize the 3D WSM to $N$ quasi-2D layers. 
Each such layer is characterized by its own density $n=n_{3D}L/N$ and its own quasi-2D intra-layer resistivity $\rho_0=\rho_{3D}N/L$. We denote by $N_d$ and $N_s$ the number of layers within a distance $d$ and $d_s$ from one of the surfaces, respectively.
Each layer within a distance $d$ from the top/bottom surface is coupled the top/bottom graphene layer according to \refEq{stand_drag}.
In the presence of an external magnetic field, each layer within a distance $d_s$ from one of the surfaces is coupled to all the other layers within a distance $d_s$ from the opposite surface via the quasi-2D non-local resistivity which we denote by $\rho_w$.    

In appendix \ref{Hall_app} we show that the effect of the intra-layer Hall effect on the non-local drag is negligible. Hence, we ignore the Lorentz force in \refEq{stand_drag}.
Without the lost of generality, we assume that $N_s\leq N_d$. Similar to the main text, we introduce the dimensionless density ratio $\xi=\frac{n_{3D}L}{n_g}$, the dimensionless non-local parameters $\Gamma=\frac{\rho_0}{2\rho_0+\rho_w}$ and $\nu=\frac{\rho_0}{\rho_0+\rho_w}$. 

The steady state equations for the longitudinal electric fields and current densities in the two graphene layers and the $N$ layers of the WSM are:
\begin{align} \label{trans_Eq}
&E_{g1}=(\rho_{g}+N_d\frac{n}{n_g}\rho_d)j_{g_1}-\rho_d\sum\limits_{i=1}^{Nd}j_i \\ \nonumber
&E_{k_1}=\left(\rho_{0}\gamma+\frac{N}{\xi}\rho_d\right)j_{k_1}-\rho_d j_0+\frac{\Gamma\rho_0}{N_g}\sum\limits_{i=N-N_g}^{N}j_i   \\ \nonumber
&E_{k_2}=\left(\rho_{0}+\frac{N}{\xi}\rho_d\right)j_{k_2}-\rho_d j_0  \\ \nonumber
&E_{k_3}=\rho_{0}j_{k_3}  \\ \nonumber
&E_{k_4}=\left(\rho_{0}\gamma+\frac{N}{\xi}\rho_d\right)j_{k_4}-\rho_d j_0+\frac{\Gamma\rho_0}{N_g}\sum\limits_{i=1}^{N_g}j_i   \\ \nonumber
&E_{g2}=(\rho_{g}+N_d\frac{n}{n_g}\rho_d)j_{g_2}-\rho_d\sum\limits_{i=N-N_d}^{N}j_i \\  \nonumber
\end{align} 
where $k_1\in [1,\, N_g]$, $k_2\in [N_g+1,\, N_d] \vee [N-N_d+1,\,N-N_g]$, $k_3\in [1+N_d,\, N-N_d]$, $k_4\in [N-N_g+1,\, N]$ and $\gamma=1-\Gamma(1+\nu)$.
\refEq{trans_Eq} may be recast in a matrix form (discrete Ohm's law):
\begin{equation}  \label{Matrix}
\mathcal{E}=\hat{\rho}\,\mathcal{J}
\end{equation}
where we defined $\mathcal{E}=\left(\begin{matrix} E_{g_1},\, E_1,\, \dotso E_N,\, E_{g_2} \end{matrix}\right)^T$, $\mathcal{J}=\left(\begin{matrix} j_{g_1},\, j_1,\, \dotso j_N,\, j_{g_2} \end{matrix}\right)^T$ and $\hat{\rho}$ is the $(N+2)\times(N+2)$ resistivity matrix of the system. In order to solve \refEq{Matrix} we must specify $N+2$ of the above quantities. In the proposed setup the current on the upper graphene layer is fixed, $j_{g_1}$. Additionally, if the lower graphene layer and the WSM are in an open circuit configuration, then $j_{g_2}=0$ and $\sum\limits_{i=1}^Nj_i=0$. The last condition automatically satisfies $\nabla\cdot \textbf{j}=0$ inside the Weyl conductor. Finally, we must demand that $\nabla\times \textbf{E}=0$ inside the WSM. This translates to $E_1=E_2=\dotso=E_N$.
Forcing these constraints, \refEq{Matrix} may be solved uniquely and all the physical and meaningful quantities have a well-defined $N\to\infty$ limit.

\section{Including the intra-layer Hall conductivity} \label{Hall_app}
Semi-classically, the local intra-layer conductivities are given by $\sigma_{xx}=\frac{\sigma_0}{1+\phi^2}$ and    
$\sigma_{xy}=\frac{\phi\sigma_0}{1+\phi^2}$. Within the framework of Drude $\phi=\frac{\sigma_{xy}}{\sigma_{xx}}=\omega_c\tau$, however, here we treat $\phi$ as a tuning parameter that captures the strength of the Hall effect. 

It is instructive to construct the full resistivity tensor for two layers in the presence of a non-local conductivity and a Hall conductivity. The full conductivity matrix is given by:
\begin{align} 
\hat{\sigma}&=\left(\begin{matrix}
\sigma_{xx}+\sigma_W& -\sigma_{xy}& -\sigma_W& 0\\
\sigma_{xy}& \sigma_{xx}+\sigma_W& 0& -\sigma_W\\
-\sigma_W& 0& \sigma_{xx}+\sigma_W& -\sigma_{xy}\\
0& -\sigma_W& \sigma_{xy}& \sigma_{xx}+\sigma_W 
\end{matrix}\right)\\ \nonumber
&=
\left(\begin{matrix}
\frac{\sigma_0}{1+\phi^2}+\sigma_W& -\frac{\phi\sigma_0}{1+\phi^2}& -\sigma_W& 0\\
\frac{\phi\sigma_0}{1+\phi^2}& \frac{\sigma_0}{1+\phi^2}+\sigma_W& 0& -\sigma_W\\
-\sigma_W& 0& \frac{\sigma_0}{1+\phi^2}+\sigma_W& -\frac{\phi\sigma_0}{1+\phi^2}\\
0& -\sigma_W& \frac{\phi\sigma_0}{1+\phi^2}& \frac{\sigma_0}{1+\phi^2}+\sigma_W 
\end{matrix}\right)
\end{align} 

\begin{figure}[t!]
\centering
{\includegraphics[width =\linewidth, angle =0] {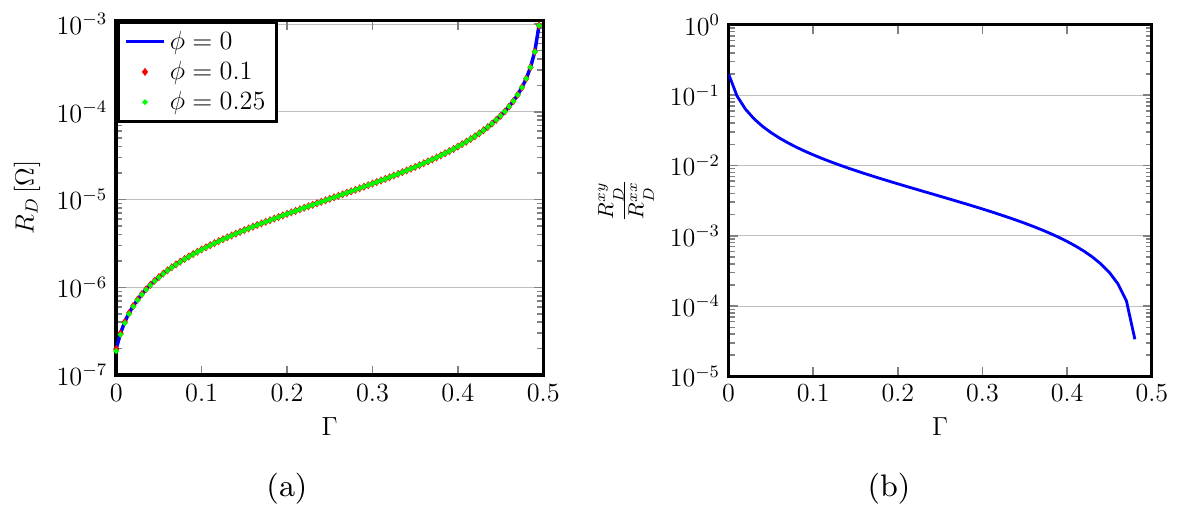}}
\caption{(a) $R^{xx}_D$ as a function of $\Gamma$ for three different values of $\phi$. (b) The ratio $\Phi$ as a function of $\Gamma$ for $\phi=0.2$. Both in (a) and (b), $L=100\,\mbox{nm}$, $d_s=d=1\,\mbox{nm}$, $r_0=5\,\Omega$, $\rho_d= 0.1\,\Omega$ and $\xi=60$. For $\Gamma\ll 1$, both $R^{xy}_D$ and $R^{xx}_D$ are small. As $\Gamma$ increases towards half, $R^{xx}_D$ increases dramatically while $R^{xy}_D$ remain almost unchanged. \label{RD_G_Hall}}
\end{figure}

Inverting this matrix yields the following resistivities:
\begin{align} 
\rho_{xx}&=\frac{1}{\sigma_0}\frac{1-\Gamma(1-\phi^2)}{1+(2\Gamma\phi)^2}\\
\rho_{xy}&=\frac{\phi}{\sigma_0}\frac{1-2\Gamma+2\Gamma^2(1+\phi^2)}{1+(2\Gamma\phi)^2}\\
\rho^{NL}_{xx}&=\frac{\Gamma}{\sigma_0}\frac{1-\phi^2(1-4\Gamma)}{1+(2\Gamma\phi)^2}\\
\rho^{NL}_{xy}&=\frac{2\Gamma\phi}{\sigma_0}\frac{1-\Gamma(1-\phi^2)}{1+(2\Gamma\phi)^2}
\end{align}  
For $\phi=0$ we get:
\begin{align} 
\rho_{xx}&=\frac{1}{\sigma_0}(1-\Gamma),\,\,\rho_{xy}=0\\ \nonumber
\rho^{NL}_{xx}&=\frac{\Gamma}{\sigma_0},\,\,\rho^{NL}_{xy}=0 \nonumber
\end{align} 
as we got in the sections above. For $\Gamma=0$, we get:
\begin{align} 
\rho_{xx}&=\frac{1}{\sigma_0},\,\,\rho_{xy}=\frac{\phi}{\sigma_0}\\ \nonumber
\rho^{NL}_{xx}&=0,\,\,\rho^{NL}_{xy}=0 \nonumber
\end{align} 
as expected. This resistivity tensor can be generalized to $N$ layers.
Including the effect of drag as before, we get a similar relation to \ref{Matrix},
\begin{equation} 
\left(\begin{matrix}
E_{g1x} \\
E_{g1y} \\
E_{1x} \\
E_{1y} \\
\vdots \\
E_{Nx} \\
E_{Ny} \\
E_{g2x} \\
E_{g2y} 
\end{matrix}\right)=\hat{\rho}\left(\begin{matrix}
j_{g1x} \\
j_{g1y} \\
j_{1x} \\
j_{1y} \\
\vdots \\
j_{Nx} \\
j_{Ny} \\
j_{g2x}\\
j_{g2y} 
\end{matrix}\right) \label{OhmHall}
\end{equation}
where now $\hat{\rho}$ is the $2(N+2)\times2(N+2)$ resistivity matrix of the system. As before, in order to solve \refEq{OhmHall} we now must specify $2(N+2)$ of the above quantities. The current on the upper graphene layer is fixed, $j_{g1x}$. Additionally, if the lower graphene layer and the WSM are in an open circuit configuration, then $j_{g_2x}=j_{g_2y}=j_{g_1y}=0$ and $\sum\limits_{i=1}^Nj_{ix}=0$ and $\sum\limits_{i=1}^Nj_{iy}=0$. The last two conditions give $\nabla\cdot \textbf{j}=0$ in the Weyl conductor. Finally, we must demand that inside the WSM $\nabla\times \textbf{E}=0$. This translate to $E_{1x}=E_{2x}=\dotso=E_{Nx}\,$ and $E_{1y}=E_{2y}=\dotso=E_{Ny}$.
With these constraints, \refEq{OhmHall} may be solved uniquely. 
The longitudinal non-local drag coefficient for a given $\phi$ may be related to the coefficient at $\phi=0$:
\begin{equation} 
R^{xx}_D[\Gamma,\phi]=R^{xx}_D[\Gamma,\phi=0]\frac{1+\eta\phi^2}{1+\phi^2}
\end{equation}  
where $R^{xx}_D[\Gamma,\phi=0]$ is given in \refEq{NLD} and $\eta=\frac{L}{d}\frac{\Gamma}{1+\left(\frac{L}{d}\right)\Gamma}$.
For $\Gamma \neq 0$ and $\frac{L}{d}\gg1$, we find that $\eta\to 1$. Therefore, $R^{xx}_D[\Gamma,\phi]\to R^{xx}_D[\Gamma,\phi=0]$.
We solved \refEq{OhmHall} numerically for specific parameters and for different values of $\phi$. See \refFig{RD_G_Hall}a. As evident, the longitudinal non-local drag resistivity is indifferent to the intra-layer Hall effect.   	

For non-zero $\phi$, there is also a non-local transverse drag coefficient.
The ratio between the longitudinal and the transverse drag coefficients is given by:
\begin{equation} 
\frac{R^{xy}_D[\Gamma,\phi]}{R^{xx}_D[\Gamma,\phi]}\equiv\Phi=\frac{\phi}{1+\eta\phi^2}\frac{1-2\Gamma}{1+\left(\frac{L}{d}\right)\Gamma}
\end{equation}
For $\Gamma=0$ we find that $\Phi=\phi$, as expected. For $\Gamma \neq 0$ and $\frac{L}{d}\gg1$ we find that $\Phi=\frac{\phi}{1+\phi^2}\frac{1-2\Gamma}{1+\left(\frac{L}{d}\right)\Gamma}$. This ratio is plotted in \refFig{RD_G_Hall}b.

Overall, there is a non-zero $R^{xy}_D$, however, it is parametrically smaller than $R^{xx}_D$. For $\Gamma\ll 1$, both $R^{xy}_D$ and $R^{xx}_D$ are small. As $\Gamma$ increases towards half, $R^{xx}_D$ increases dramatically while $R^{xy}_D$ remain almost unchanged. 

\section{Evaluation of the local drag resistance} \label{drag_app}
In order to evaluate the local drag resistivity, $\rho_d$, we treat the drag layer of the WSM (up to a distance of $d$ from one of surfaces) as an effective 2D layer with a known density $n_d=n_{3D}d$. We also assume that both graphene layers are doped away from the neutrality point, and have a finite density $n_g$. 

Given two layers with densities $n_g$ and $n_d$ which are separated by a distance $\ell$, the drag resistivity may be obtained from the standard Kubo
formula approach within the diagrammatic perturbation theory \cite{DragOreg,DragKarsten} or from a memory function formalism \cite{DragMec}:
\begin{equation} \label{drag_form}
\rho_d=\frac{\hbar}{16e^2\pi n_d n_g T}\int{\frac{d^2\textbf{q}d\omega q^2}{(2\pi)^2}\frac{\left|U_{12}(\textbf{q},\omega)\right|^2F_1(\textbf{q},\omega)F_2(\textbf{q},\omega)}{\sinh^2{\left(\frac{\hbar\omega}{2T}\right)}}}
\end{equation} 
where

\begin{align} 
&F_i(\textbf{q},\omega)=\Im\left[\Pi_i(\textbf{q},\omega)\right]=\\ \nonumber
&\int{\frac{d^2\textbf{k}}{(2\pi)^2}\left(f_{FD}(\epsilon_\textbf{k})-f_{FD}(\epsilon_{\textbf{k}+\textbf{q}})\right)\delta\left(\epsilon_{\textbf{k}+\textbf{q}}-\epsilon_{\textbf{k}}-\hbar\omega\right)}
\end{align} 
is the imaginary part of the is the single-layer retarded polarization operator and
\begin{equation} 
U_{12}^{-1}(\textbf{q},\omega)=e^{q\ell}\left(\frac{q}{2\pi e^2}+\Pi_1+\Pi_2\right)+ \Pi_1\Pi_2\frac{4\pi e^2}{q}\sinh{(q\ell)}
\end{equation}
is the screened inter-layer Coulomb interaction, where again $\Pi_i=\Pi_i(\textbf{q},\omega)$ is the single-layer retarded polarization operator.

For two-dimensional, non-interacting electron gas in the ballistic regime the $F$ function is given by:
\begin{equation} \label{F_func}
F_i(\textbf{q},\omega)=\nu_i\frac{\omega}{v_{f,i}q}\Theta(v_{f,i}q-\omega)
\end{equation}
where $\nu_i$ and $v_{f,i}$ are the density of states and the Fermi velocity in the $i$'s layer, respectively.

Additionally, in the limit of strong screening, the screened inter-layer interaction can approximated by:
\begin{equation} \label{int}
U_{12}(\textbf{q},\omega)\approx \frac{\pi e^2}{q_{_{TF,1}}q_{_{TF,2}}}\frac{q}{\sinh{(q\ell)}}
\end{equation}
where $q_{_{TF,i}}=2\pi e^2\nu_i$ is the inverse Thomas-Fermi screening length. 

Employing \refEq{F_func} and \refEq{int}, keeping in mind the degeneracies in each layer, the integrals in \refEq{drag_form} may be evaluated,
\begin{equation} \label{drag_close}
\rho_d=\frac{h}{e^2}\frac{\zeta(3)}{256\pi^2}\frac{T^2}{e^4\ell^4 (n_g n_d)^{3/2}}
\end{equation} 

Denoting $\mathcal{N}=(n_g n_d)^{1/2}$, the local drag resistivity may be written as:
\begin{equation}
\rho_d\sim4.43\cdot 10^{6}\frac{T^2}{\mathcal{N}^3\ell^4}\,[\Omega]
\end{equation} 
where $\mathcal{N}$ is measured in $\mbox{cm}^{-2}$, $\ell$ in $\mbox{cm}$ and $T$ in Kelvin. Finally, we may define the following typical dimensionless parameters (of the order of unity):

\begin{equation} 
\Delta=\frac{\mathcal{N}}{5\cdot 10^{11}\mbox{cm}^{-2}},\,t=\frac{T}{100\mbox{K}} \mbox{ and } \chi=\frac{\ell}{10\mbox{nm}}
\end{equation} 
which yields:

\begin{equation}
\rho_d\sim 0.35\frac{t^2}{\Delta^3\chi^4}\,[\Omega]
\end{equation}

\end{document}